\documentclass[fleqn,twoside]{article}
\usepackage{espcrc2}

\usepackage{graphicx}
\usepackage[figuresright]{rotating}

\newcommand{\AmS}{{\protect\the\textfont2
  A\kern-.1667em\lower.5ex\hbox{M}\kern-.125emS}}

\title{$J/\Psi$ production in two-photon collisions at next-to-leading order}

\author{L. Mihaila\address[MCSD]{II. Institut f\"ur Theoretische Physik,Universit\"at Hamburg,\\ 
        Luruper Chaussee 149, 22761 Hamburg, Germany}%
        \thanks{Talk given at International Conference and 
           Loops and Legs in Quantum Field 
           Theory, Zinnowitz, Germany, 25-30 April 2004.}
           }

\begin{document}

\begin{abstract}
In this paper, we report on the calculation of the cross section of $J/\Psi$
plus jet inclusive production in direct $\gamma\gamma$ collisions 
at next-to-leading order within the factorization formalism of nonrelativistic
quantum chromodynamics (NRQCD). Theoretical predictions for the future
$e^+e^-$ linear collider TESLA are also presented.   
\vspace{1pc}
\end{abstract}

\maketitle

\section{INTRODUCTION}
Heavy quarkonium played an important role in establishing the asymptotic
freedom of QCD, as its mass is much larger than
the low-energy scale of the strong interactions $\Lambda_{QCD}$. On the other hand, the
compensation of colour in heavy quarkonium gives new insights into the
confinement mechanism, a nonperturbative process which is still not fully
understood.\\
The factorization approach based on the nonrelativistic QCD (NRQCD) \cite{cas,bbl}
provides a rigoros and systematic  framework for calculating
 perturbative and relativistic corrections to heavy quarkonium
production and annihilation rates. This formalism implies a separation of
short-distance coefficients, which can be expressed as perturbative expansions
in the strong-coupling constant $\alpha_s$, from the long-distance matrix
elements (MEs), which must be extracted from the experiment. The MEs are
predicted to scale with a definite power of the heavy quark relative velocity
$v$. In this way, the theoretical calculations are organized as double
expansions in $\alpha_s$ and $v$. \\
Within the NRQCD formalism, the colour-octet (CO)
processes contribute to the production and decay rates at some level. Their
quantitative significance was assessed for inclusive charmonium hadroproduction 
observed at the Fermilab Tevatron \cite{abe}. However, it is necessary to test the
universality of the MEs in other kinds of production processes, such as
$\gamma\gamma$ collisions. This process was studied at LEP2, where the photons
originated from hard initial-state bremsstrahlung. Hopefully,  the future
$e^{+}e^{-}$ linear collider will be built, for which an additional source of hard photons would
be provided by beamstrahlung.\\
However, the leading-order theoretical predictions to be
compared with the experimental data suffer from considerable uncertainties,
mostly from the dependences on the renormalization and factorization scales and
from the lack of information on the nonperturbative MEs. In this paper, we
report on the calculation of the next-to-leading order (NLO) corrections
 to the inclusive production of $J/\Psi$ mesons in $\gamma\gamma$ collisions \cite{kkms}. 
This is a first step in a comprehensive programme concerning the study of charmonium 
production at NLO within the NRQCD framework.

\section{NLO CORRECTIONS TO $J/\Psi$ \\DIRECT PHOTOPRODUCTION}

In the following, we take into consideration the process $\gamma\gamma\to J/\Psi +j+X$,
where X denotes the hadronic remnant possibly including a second jet. We
restrict the analysis to $j$ and $X$ free of charm quarks and to finite values
for the $J/\Psi$ transverse momentum $p_T$. Two-photon processes can be
modelled by assuming that each photon either interacts directly (direct
photoproduction) or fluctuates into hadronic components (resolved
photoproduction). Thus, the above mentioned process receives contributions
from the direct, single-resolved, and double-resolved channels. Furthermore,
we concentrate on the NLO calculation of the inclusive cross section for direct
photoproduction.\\
At LO in $\alpha_s$, there is only one partonic subprocess that scales like $v^2$, namely
\begin{equation}
\gamma\gamma\to c\bar{c}[{}^3\!S_1^{(8)}]+g.
\label{eq:born}
\end{equation}
The representative Feynman diagrams together with the analytic expression 
of the LO cross section for the process (\ref{eq:born}) can be found  in Ref.~\cite{npb}.\\
In the calculation of the production rates beyond LO in $\alpha_s$,
ultraviolet (UV), infrared (IR), and Coulomb divergences arise and need to be
regularized. One of the most convenient method to handle the UV and IR
singularities which are present in the short-distance
coefficients as well as in the MEs is dimensional regularization. We introduce a 't Hooft
mass $\mu$ and a factorization mass $M$ as unphysical scales, and formally distinguish
between UV and IR poles.
The Coulomb singularities are regularized by assigning a
small relative velocity between the $c$ and $\bar{c}$ quarks.\\
 For the
extraction of the short-distance coefficients we apply the projector formalism
of Refs.~\cite{kap,gre}.\\ 
In the following, we study  virtual and real radiative corrections separately, 
as their calculation  requires different approaches.

\subsection{Virtual Corrections}
\begin{figure}
\includegraphics[bbllx=192pt,bblly=672pt,bburx=407pt,bbury=735pt]{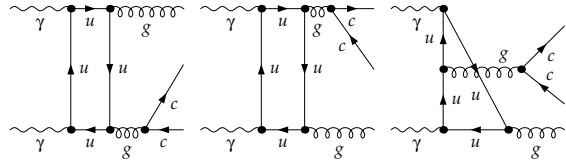}
\vspace*{-1cm}
\caption{Light-quark box diagrams contributing to the virtual corrections to
  $\gamma\gamma\to c\bar{c}[{}^3\!S_1^{(8)}]+g$.}
\label{fig:qq}
\vspace*{-3mm}
\end{figure}

Feynman diagrams that generate the virtual corrections to the process
(\ref{eq:born}) can be collected into two classes. The first one is obtained by
attaching one virtual gluon in all possible ways to the tree-level
diagrams. They include self-energy, triangle, box, and pentagon diagrams. Loop
insertions in external gluon or $c$-quark lines are taken into consideration
in the respective wave-function renormalization constants. The self-energy and
triangle diagrams are in general UV divergent; the triangle, box and pentagon
diagrams are in general IR divergent. The pentagon diagrams comprising only
abelian gluon vertices also contain Coulomb singularities which  
cancel out similar poles in the radiative corrections to the operator
$\left\langle{\cal O}^H\left[{}^3\!S_1^{(8)}\right]\right\rangle$. For the
analytical treatment of the abelian five-point functions, we refer to 
Ref.~\cite{Beenakker:2002nc}.\\
The diagrams of the second class comprise light-quark loops. The individual contributions
arising from the  triangle diagrams are equal to zero according to Furry's
theorem \cite{fur}. The box diagrams, selectively shown in  Fig.~\ref{fig:qq},
 contain UV and IR singularities, but
their sum is finite.\\
The UV divergences comprised in self-energy and triangle diagrams are
cancelled upon renormalization of the QCD gauge coupling
$g_s=\sqrt{4\pi\alpha_s}$, the charm-quark mass $m$ and field $\Psi$, and the
gluon field $A$. We adopt the on-mass-shell (OS) scheme to renormalize $m$,
 $\Psi$, and $A$, while for $g_s$ we employed the modified minimal-subtraction
 ($\overline{\rm MS}$) scheme.
\subsection{Operator Renormalization}
For a consistent NLO analysis, one should also  consider
higher-order corrections in $\alpha_s$ to the four-quark operator
$\left\langle{\cal O}^H\left[{}^3\!S_1^{(8)}\right]\right\rangle$ within
NRQCD. This bare $d$-dimensional operator has mass dimension $d-1$ while its
renormalized version, which is extracted from the experimental data, has mass
dimension $3$. We thus introduce the 't Hooft mass scale of NRQCD, $\lambda$,
to compensate for the difference between bare and renormalized operators.  The
corresponding tree-level and one-loop diagrams are depicted in Fig.~5 of
Ref.~\cite{gre}. Using dimensional regularization, the NRQCD Feynman rules 
in the quarkonium rest frame, and
firstly expanding the integrands as Taylor series in $1/m$, and
afterwards performing the integration over the loop momentum, we obtain the
unrenormalized one-loop result
\begin{eqnarray}
&&\left\langle{\cal O}^H\left[{}^3\!S_1^{(8)}\right]\right\rangle_1
=\left\langle{\cal O}^H\left[{}^3\!S_1^{(8)}\right]\right\rangle_0\nonumber\\
&\times&\left[1+\left(C_F-\frac{C_A}{2}\right)\frac{\pi\alpha_s}{2v}\right]\nonumber\\
&+&\frac{4\alpha_s}{3\pi m^2}\left(\frac{4\pi\mu^2}{\lambda^2}\right)^\epsilon
\exp(-\epsilon\gamma_E)\left(\frac{1}{\epsilon_{\rm UV}}-\frac{1}{\epsilon_{\rm IR}}\right)
\nonumber\\
&\times&
\bigg(\sum_{J=0}^2
C_F\left\langle{\cal O}^H\left[{}^3\!P_J^{(1)}\right]\right\rangle \nonumber\\
&+&{} \sum_{J=0}^2 B_F\left\langle{\cal O}^H\left[{}^3\!P_J^{(8)}\right]\right\rangle\bigg),
\label{eq:reg}
\end{eqnarray}
where the subscript 0 labels the tree-level quantity and $\mu$ is the 't~Hooft
mass scale of QCD.\\
The presence of UV divergences indicates that the 
$\left\langle{\cal O}^H\left[{}^3\!S_1^{(8)}\right]\right\rangle$ operator
needs renormalization. For this, we choose the $\overline{\rm MS}$  scheme  so that
the $1/\epsilon_{UV}$ pole in Eq.~(\ref{eq:reg}) is cancelled out by the
$1/\epsilon_{UV} +ln(4\pi)-\gamma_E$ pole in the coefficient of the
counterterm ME. The term proportional with $1/v$ represents the Coulomb
singularity, which compensates  a similar term in the virtual
corrections. After the renormalization of the NRQCD ME and cancellation of the
Coulomb singularity,  an IR counterterm at ${\cal O}(\alpha_s)$ is generated
from Eq.~(\ref{eq:reg}),
which is indispensable to render the overall NLO result finite. This feature
will be discussed in some detail in the next section.
\subsection{Real Corrections}
The real corrections to the process   (\ref{eq:born}) arise from the partonic
subprocesses
\begin{equation}
\gamma(k_1)+\gamma(k_2)\to c\overline{c}[n](p)+g(k_3)+g(k_4),
\label{eq:ccgg}
\end{equation}
where $n={}^3\!P_J^{(1)},{}^1\!S_0^{(1)},{}^1\!S_0^{(8)},{}^3\!S_1^{(8)},{}^3\!P_J^{(8)}$, and
\begin{equation}
\gamma(k_1)+\gamma(k_2)\to c\overline{c}[n](p)+q(k_3)+\overline{q}(k_4),
\label{eq:ccqq}
\end{equation}
where $n={}^1\!S_0^{(8)},{}^3\!S_1^{(8)},{}^3\!P_J^{(8)}$. For the respective
Feynman diagrams we refer to Fig.~2 and Fig.~3 of Ref.~\cite{npb}. Note that
the colour-singlet states $n={}^3\!S_1^{(1)}$ in process (\ref{eq:ccgg})
and $n={}^3\!P_J^{(1)},{}^1\!S_0^{(1)}$ in  process (\ref{eq:ccqq})  are forbidden 
by the Furry's theorem \cite{fur} and  colour conservation, respectively.

Integrating the squared MEs of the processes (\ref{eq:ccgg}) and
(\ref{eq:ccqq}) over the three-particle phase space while
keeping the value of $p_T$ finite, we encounter IR singularities, which can be
of the soft and/or collinear type.
In order to systematically extract these singularities, it is useful to slice 
the phase space by introducing infinitesimal
dimensionless cut-off parameters $\delta_i$ and $\delta_f$, which are
connected with the initial and final states, respectively \cite{harris}.
In the case of process (\ref{eq:ccgg}) we distinguish  {\it soft}, {\it
  final-state collinear} and {\it hard} regions of the phase space. In case of
process (\ref{eq:ccgg}) we differentiate {\it initial-state collinear} and
{\it hard} regions of the phase space. While the individual contributions
depend on the cut-off parameters $\delta_i$ or $\delta_f$, their sum must be
independent of them. We used the numerical verification of the cut-off
independence as a check for our calculation.

The contributions originating from process (\ref{eq:ccgg}) with
$n={}^3\!S_1^{(8)}$ contain IR (soft and final-state collinear) singularities
which cancel out singularities of the same type comprised in the virtual
corrections, as required by the Kinoshita-Lee-Nauenberg theorem \cite{kln}.
The subprocesses with $n={}^3\!P_J^{(1)},{}^3\!P_J^{(8)}$ generates
contributions containing only soft divergences. These are cancelled out by
the IR poles leftover in the  renormalized 
$\left\langle{\cal O}^H\left[{}^3\!S_1^{(8)}\right]\right\rangle$ operator,
which organise themselves as coefficients of the 
$\left\langle{\cal O}^H\left[{}^3\!P_J^{(1,8)}\right]\right\rangle$
operators.\\
According to the mass factorization theorem \cite{dew}, the form of the
collinear singularities present in the squared amplitudes of 
processes (\ref{eq:ccqq}), which are related with  an
incoming-photon leg, is universal and the pole can be absorbed into the bare
PDF of the antiquark $\bar{q}$ inside the resolved photon. As a consequence,
the real MEs acquire an explicit dependence on the mass factorization scale $M$.
 In turn, the
resolved-photon contribution is evaluated with the renormalized photon PDF,
which are also $M$ dependent. In the sum of these two contributions, the $M$
dependence cancels up to terms beyond NLO. 
\section{PHENOMENOLOGICAL STUDY}

\begin{figure}
\includegraphics[scale=0.4]{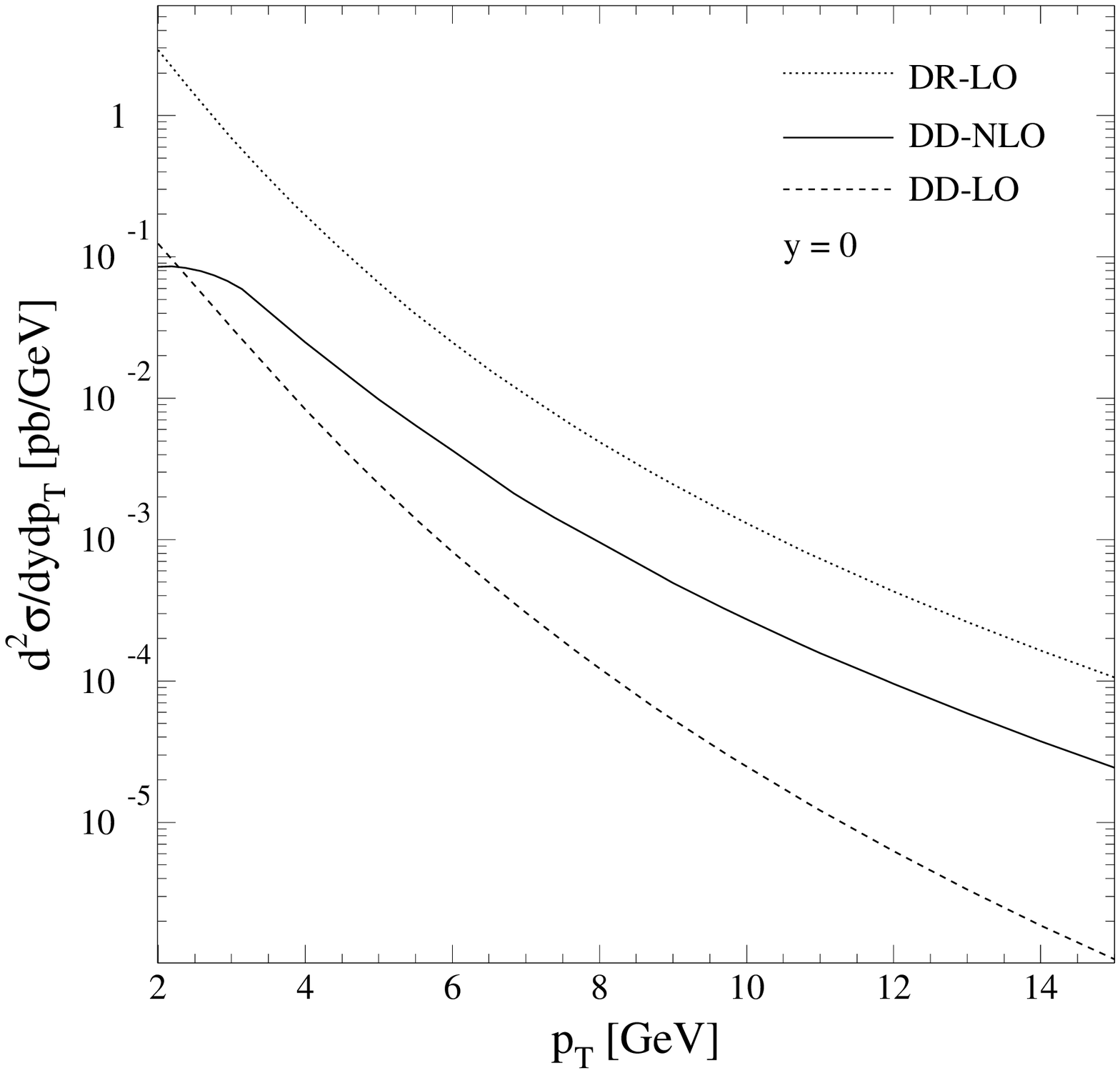}
\hspace*{4,1cm}(a)\\
\includegraphics[scale=0.4]{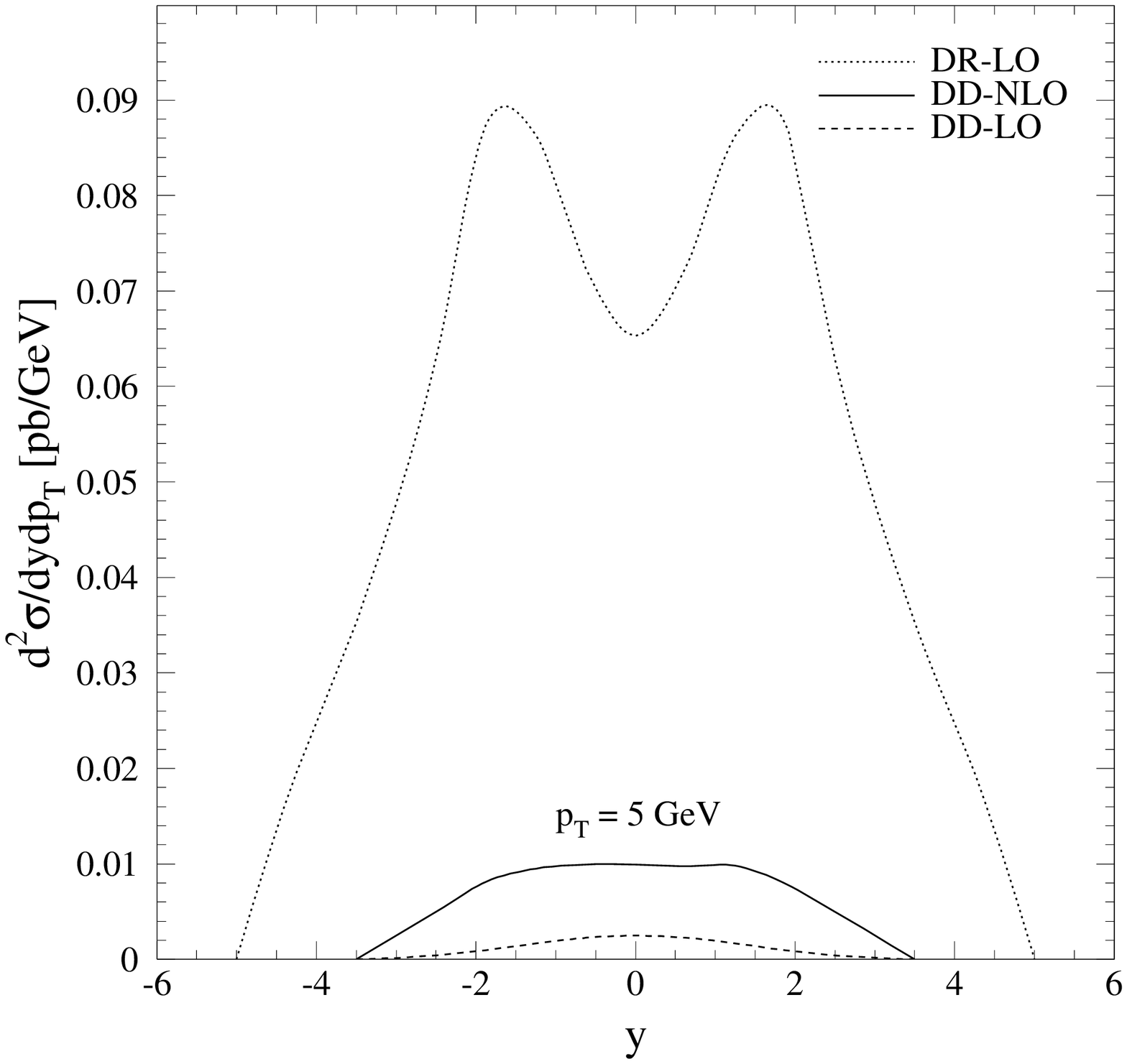}
\hspace*{4,1cm}(b)
\caption{Differential cross section $d^2\sigma/dp_T\,dy$ of
$e^+e^-\to e^+e^-J/\psi+X$ at TESLA with $\sqrt s=500$~GeV
 (a) for  $y=0$ as a function of $p_T$ and
(b) for $p_T=5$~GeV as a function $y$.}
\label{fig:xs}
\end{figure}

In the following, we study the phenomenological significance of the process\\
$\gamma\gamma\to J/\Psi +j +X$
at  TESLA in its $e^+e^-$ mode, with $\sqrt s=500$~GeV and the
photon sources bremsstrahlung and beamstrahlung  coherently superposed.   
For the energy spectrum of the  photons we refer to
 Eq.~(27) of Ref.~\cite{fri} and  
Eq.~(2.14) of Ref.~\cite{pes}, and for the designed experimental parameters to
Ref.~\cite{theta} and Ref.~\cite{tesla}, respectively.

We use $m=1.5$~GeV, $\alpha=1/137.036$, and the two-loop formula for 
$\alpha_s^{(n_f)}(\mu)$ \cite{pdg} with $n_f=3$ active quark
flavours and $\Lambda_{\rm QCD}^{(3)}=299$~MeV  \cite{grs}. As for photon
PDFs, we employ the NLO set from Glu\"{u}ck, Reya, and Schienbein (GRS)
\cite{grs}, which  are implemented in the
fixed fixed-flavour-number scheme, with $n_f=3$. Our default choice of
renormalization and factorization scales is $\mu=M=m_T$ and $\lambda=m$.
As for  $J/\psi$, $\chi_{cJ}$, and $\psi^\prime$ MEs, we adopt the 
LO sets determined in Ref.~\cite{bkl} using the LO set of proton PDFs from
Martin, Roberts, Stirling, and Thorne (MRST98LO) \cite{mrst}.

In Fig.~\ref{fig:xs}, we study $d^2\sigma/dp_T\,dy$ (a) for $y=0$ as a
function of $p_T$ and (b) for $p_T=5$~GeV as a function of $y$. The solid  and
dashed lines represent NLO and LO results, respectively, for the direct
photoproduction. The dotted lines correspond to LO result of the
single-resolved photoproduction evaluated with the LO versions of the 
$\alpha_s^{(n_f)}(\mu)$ and the photon PDFs. Notice that we do not consider
$p_T$ values smaller than $2$~GeV, where additional IR singularities occur.\\
From Fig.~\ref{fig:xs}(a), we observe that, the NLO result of direct
photoproduction falls off with increasing $p_T$ more slowly than LO one. This
feature may be accounted for by the {\it fragmentation-prone} partonic
subprocesses that contribute at NLO result, while they are absent at
LO. Such subprocesses contain a gluon with a small virtuality that splits into
a $c\bar{c}$ pair in the Fock state $n={}^3\!S_1^{(8)}$. They generally
provide dominant contributions at $p_T\gg2m$ due to the
presence of a large gluon propagator. In single-resolved photoproduction, 
a fragmentation-prone partonic subprocess
already contributes at LO. This explains why the solid and dotted curves in 
Fig.~\ref{fig:xs}(a) run parallel in the upper $p_T$ range. In the lower $p_T$
range, the fragmentation-prone partonic subprocesses do not matter, as the
gluon propagator becomes finite and the Fock state $n={}^3\!S_1^{(8)}$ is
already present at LO.\\
As one can notice from Fig.~\ref{fig:xs}(b), at $p_T=5$~GeV, single-resolved 
photoproduction is still overwhelming. The two pronounced maxima of the LO
single-resolved result may be explained by the occurrence of a virtual gluon
in a $t$ channel that can be almost collinear with the incoming quark $q$ or
by the one of a virtual quark in the $u$ channel that can become almost
collinear with the incoming photon. The NLO result increases towards forward
and backward directions due to the finite reminders of the initial-state
collinear singularities that were absorbed into the photon PDFs.

\section{CONCLUSIONS}
The experimental verification of the NRQCD factorization hypothesis is of
great importance especially for charmonium, because the charm quark mass might
not be large enough to justify the nonrelativistic approximation.\\
In this paper, we studied at NLO the inclusive production of prompt $J/\Psi$
mesons with finite values of $p_T$. This is the first time that an inclusive
$2\to 2$ production process was treated at NLO within the NRQCD factorization
formalism. A complete NLO calculation of prompt $J/\Psi$ production in
$\gamma\gamma$ collisions allows a NLO treatment of photo 
and hadroproduction as well. This  will provide a solid basis for
the attempt to accommodate the NRQCD theoretical predictions with the experimental
data to be taken at high-energy colliders, such as the Tevatron (RunII) and
HERA (HERA-II). 

\bigskip
\noindent
{\bf Acknowledgements}
\smallskip

The author thanks M. Klasen, B.A. Kniehl, and M. Steinhauser for a fruitful collaboration.

\end{document}